\documentclass[aps, prb, twocolumn, superscriptaddress, longbibliography]
    {revtex4-2}

\usepackage{graphicx}
\usepackage{epstopdf}
\usepackage{dcolumn}
\usepackage{bm}
\usepackage{color}
\usepackage{hyperref}
\usepackage{natbib}
\graphicspath{{figures/}{../figures/}}

\begin{document}


\title{Energy spectrum of semimetallic HgTe quantum wells}

\author{Jan~Gospodari\v{c}}
\author{Alexey~Shuvaev}
\affiliation{Institute of Solid State Physics, Vienna University of
Technology, 1040 Vienna, Austria}
\author{Nikolai~N.~Mikhailov}
\author{Ze~D.~Kvon}
\affiliation{Rzhanov Institute of Semiconductor Physics and
Novosibirsk State University, Novosibirsk 630090, Russia}
\author{Elena G. Novik}
\affiliation{Institute of Theoretical Physics,
Technische Universit\"{a}t Dresden, 01062 Dresden, Germany}
\author{Andrei Pimenov}
\affiliation{Institute of Solid State Physics, Vienna University of
Technology, 1040 Vienna, Austria}

\begin{abstract}
Quantum wells (QWs) based on mercury telluride (HgTe) thin films provide a large scale of unusual physical properties starting from an insulator via a two-dimensional Dirac semimetal to a three-dimensional topological insulator. These properties result from the dramatic change of the QW band structure with the HgTe film thickness. Although being a key property, these energy dispersion relations cannot be reflected in experiments due to the lack of appropriate tools. Here we report an experimental and theoretical study of two HgTe quantum wells with inverted energy spectrum in which two-dimensional semimetallic states are realized. Using magneto-optical spectroscopy at sub-THz frequencies we were able to obtain information about electron and hole cyclotron masses at all relevant Fermi level positions and different charge densities. The outcome is also supported by a Shubnikov-de Haas analysis of capacitance measurements, which allows obtaining information about the degeneracy of the active modes. From these data, it is possible to reconstruct electron and hole dispersion relations. Detailed comparative analysis of the energy dispersion relations with theoretical calculations demonstrates a good agreement, reflecting even several subtle features like band splitting, the second conduction band, and the overlaps between the first conduction and first valence band.  Our study demonstrates that the cyclotron resonance experiments can be efficiently used to directly obtain the band structures of semimetallic 2D materials.


\end{abstract}

\date{\today}


\maketitle

\begin{figure*}[tbp]
	\begin{minipage}{0.65\linewidth}
		\centering
		\includegraphics[width=0.95\linewidth]{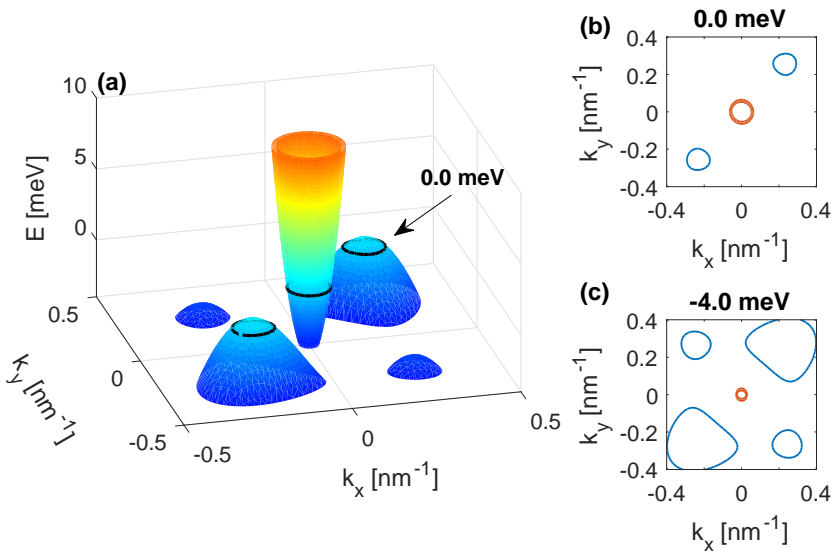}
	\end{minipage}
	\begin{minipage}{0.3\linewidth}
	\caption{\textit{Theoretical band structure for a semimetallic QW.} (a) The first valence band  and the first conduction band of the $14.1$\,nm-thick HgTe layer at the charge neutrality point with the Fermi energy of $0$~meV. (b) Cross section of the band structure at $E=0$~meV. Blue: holelike Fermi surface ($\partial A /\partial E <0$) from the islands in the $(\pm1,\pm1)$ directions. Orange: electronlike Fermi surface ($\partial A /\partial E > 0$) from the conduction band. (c) Fermi surface of the hole-doped sample with $E=-4$\,meV.}
	\label{fig_th}
	\end{minipage}
\end{figure*}

\section{Introduction}\label{Intro}

Quantum wells (QWs) based on strained HgTe films have been studied extensively in recent years revealing numerous exotic properties~\cite{kvon_ufn_2020, koenig_jpsj_2008}. These properties arise due to the shifts of the  s-like E1 and p-like H1 and H2 bands of HgTe with the QW thickness. For the thickness below critical $d_c \sim 6.5$\,nm an energy gap opens between H1 valence and E1 conduction bands and an insulating state can be realized. Both bands are touching at $d=d_c$ forming a two-dimensional (2D) semimetal with zero gap and with Dirac dispersion~\cite{bernevig_science_2006}. Further increasing of the thickness leads again to an opening of a gap. However, in this case, the original E1 and H1 bands are interchanged forming an inverted band structure and a quantum spin Hall insulator~\cite{bernevig_science_2006, konig_science_2007}.
If a bulk HgTe layer $d \gtrsim 50$\,nm is grown on a CdTe substrate, tensile strain due to lattice mismatch splits the originally degenerate light and heavy $\Gamma_8$ hole bands at the center of the Brillouin zone, forming a three-dimensional topological insulator~\cite{fu_prb_2007, brune_prl_2011}.

HgTe wells in the thickness range of about 10-30\,nm
represent unique examples of two-dimensional (2D) semimetals, where electrons and holes coexist simultaneously~\cite{Kvon2008, ortner_prb_2002}. Here the H2 valence and H1 conduction bands indirectly overlap forming a negative gap in the meV range.
With a few exceptions~\cite{minkov_physe_2020}, studies of HgTe quantum wells in a semimetallic state~\cite{Olshanetsky2009, Kozlov2011, Kvon2011, Minkov2013, Olshanetsky2012, Minkov2017} generally concluded that while the measured properties of the conduction band agree reasonably well with the theoretical models the valence band spectrum does not. The results have generally shown the valence subbands being strongly anisotropic, forming four local maxima at non-zero k-values, with an overlap of a few meV with the rotationally symmetric conduction subband. However, the mismatch between experimental data and model calculations, such as the band overlap and the hole effective mass, indicated that the theoretical approach to this problem is not fully established. Moreover, recent experiments on samples with (013) surface orientation~\cite{Minkov2017} suggested a two-fold valley degeneracy of the top valence subbands.

Here we investigate two semimetallic HgTe QWs by the analysis of the cyclotron resonance (CR). Applying the recently established technique~\cite{shuvaev_prb_2017, Gospodaric2020} allows us to directly obtain the band structures of these 2D structures. The analysis of the Shubnikov-de Haas (SdH) oscillations seen in the capacitance of the samples gave additional insight into the properties of the charge carriers in the system, providing the degeneracies of the bands. With two experimental techniques, we were able to probe the top valence subband states, the first conduction subband, and even the second conduction subband.
The experimental results of the CR analysis and magnetotransport measurements are compared with the $\mathbf{k \cdot p}$ model, showing a good overlap. The combination of the bulk-inversion asymmetry and the structure-inversion asymmetry leads to an appearance of two hole islands in the valence subband~\cite{Winkler2000, Ganichev2014}.

\section{Results and discussion}
\label{secres}

Figure~\ref{fig_th} shows the band structure of a two-dimensional semimetal obtained via  $\mathbf{k \cdot p}$ calculations as detailed in Section~\ref{secmod}. These results confirm the rotational symmetry of the conduction bands which allows reconstructing the experimental band structure via integration of the simplified Eq.\,(\ref{eqCR2}) (see Materials and Methods Section below).
The valence bands in semimetallic HgTe quantum wells show strongly anisotropic behavior~\cite{Minkov2017}, demonstrating two pronounced maxima ("islands") in the first valence subband H2. As demonstrated by Fig.\ref{fig_th}(b), at low hole concentrations, i.e. close to the band maxima, the Fermi surfaces of the islands can be reasonably well approximated by circles shifted by $k=\pm (k_0,k_0)$ from the $\Gamma$-point with an effective radius $k_{eff}$. Here, $k_0 \approx 0.25$~nm$^{-1}$ for the $14.1$\,nm thick sample and $k_0 \approx 0.22$~nm$^{-1}$ for the $22$\,nm sample. In this approximation, the approach to reconstruct the energy dispersion given in Eqs.\,(\ref{eqCR}) and (\ref{eqCR2}) is justified.


\subsection{Cyclotron resonance}
\label{secres_CR}

\begin{figure}[tbp]
	\includegraphics[width=1\linewidth]{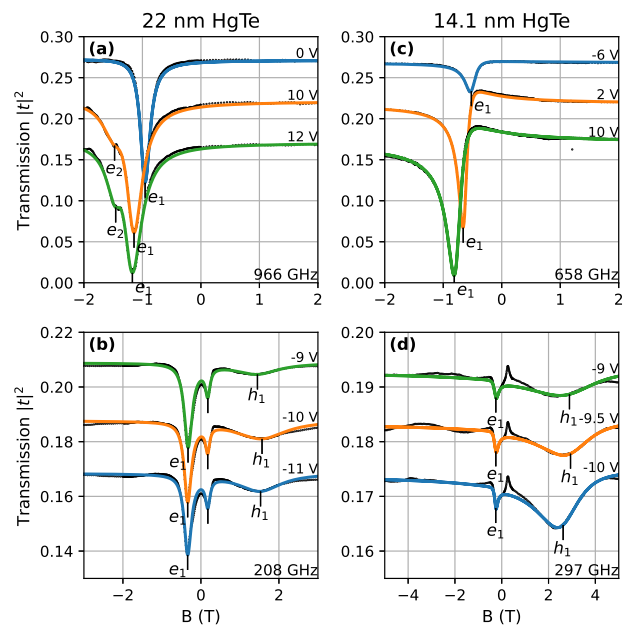}
	\caption{\textit{Cyclotron resonance with circularly-polarized light.}
		The intensity of the transmitted radiation $|t_+|^2$ through the $22$\,nm (a,b) and the $14.1$\,nm sample (c,d) as a function of the external magnetic field for fixed frequencies, as indicated. Resonance features for positive and negative fields correspond to holes and electrons, respectively. Black points - experiment, solid lines - theoretical transmission based on Drude conductivity, Eq.\,(\ref{eq_drude}). The absolute scales refer to the lowest curves, others are shifted for clarity.}
	\label{fig_spec}
\end{figure}
\begin{figure*}[tbp]
	\begin{minipage}{0.5\textwidth}%
		\centering
		\includegraphics[width=\textwidth]{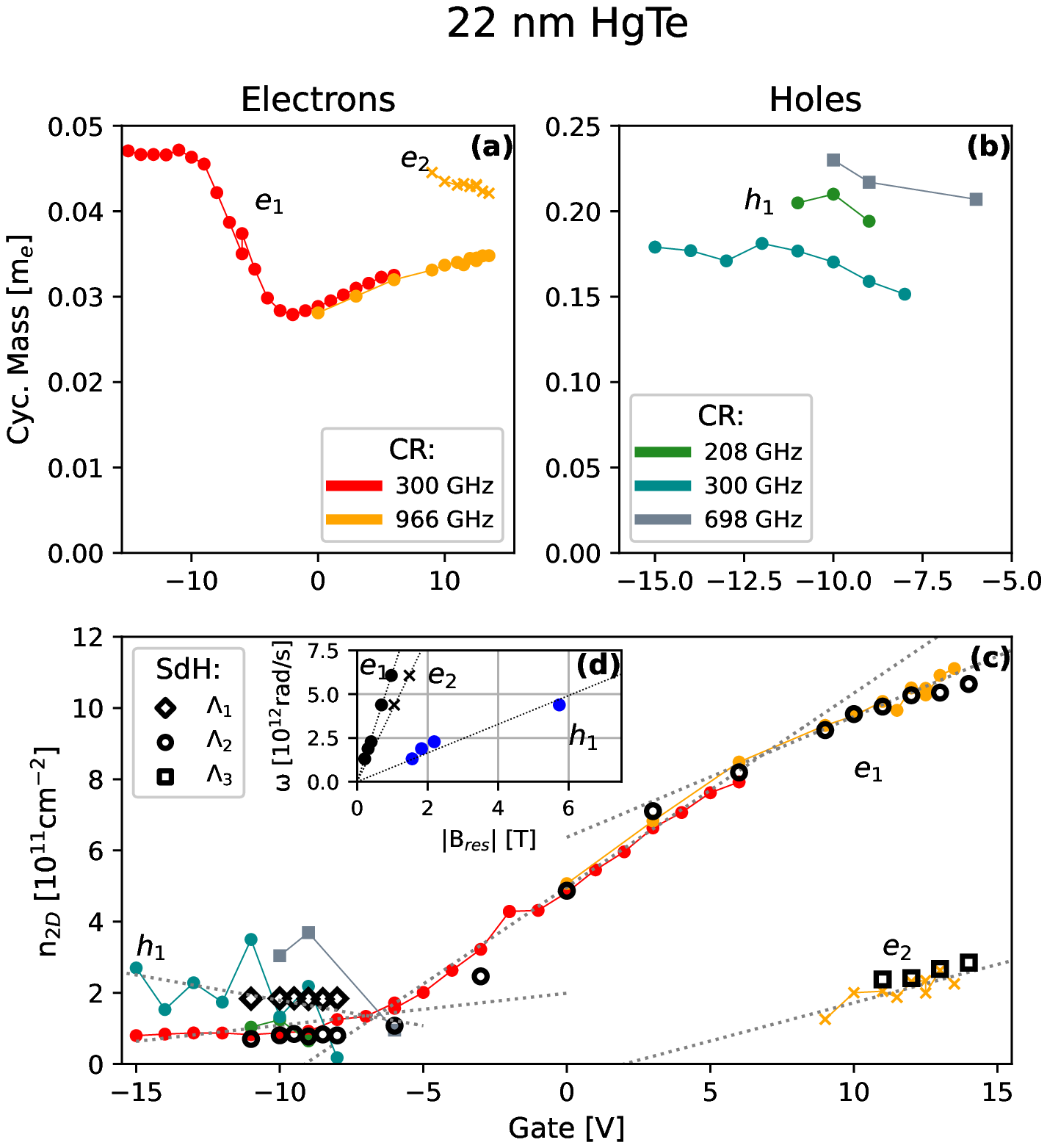}
	\end{minipage}%
	\hfill
	\begin{minipage}{0.5\textwidth}%
	\centering
		\includegraphics[width=\textwidth]{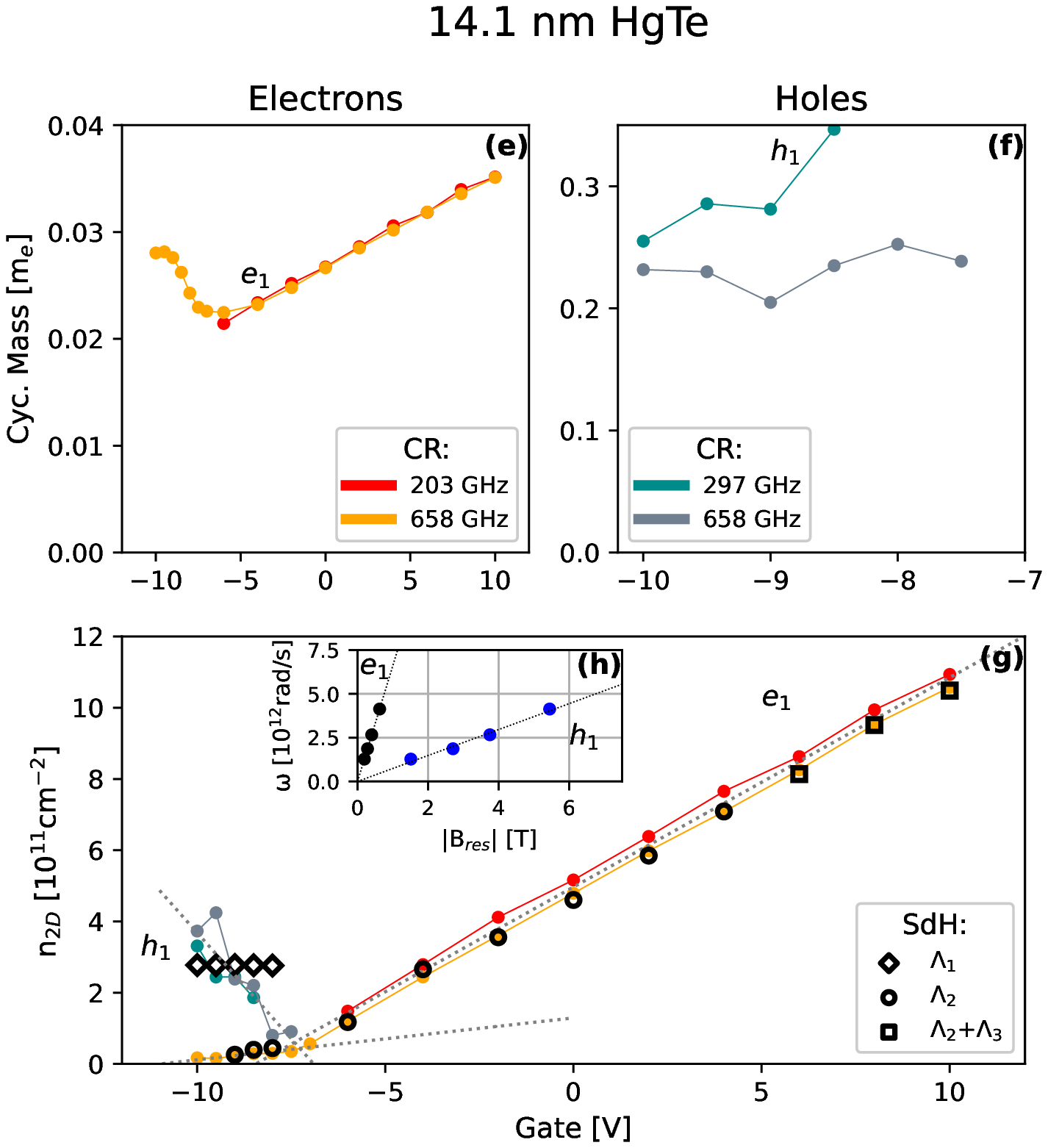}
	\end{minipage}%
	\caption{\textit{Electrodynamic parameters of the cyclotron resonances in HgTe.} (a-d) -- $22$\,nm-HgTe sample, (e-h) -- $14.1$\,nm-HgTe sample. Colored symbols are experimental data from the fits of the spectra in Fig.\,\ref{fig_spec}. The dotted lines in (c,g) serve as guides for the eye. Black empty symbols correspond to the carrier concentration resulting from the SdH analysis (see Fig.\,SI.2 in the Supplementary in Ref.~\cite{supp}). The insets (d,h) show the field dependence of the cyclotron resonance demonstrating linear behavior within the quasi-classical approximation according to Eq.\,(\ref{eqCR}).}
	
	\label{fig_par}
\end{figure*}

Figure~\ref{fig_spec} shows the spectra of the field-dependent transmission in the
geometry with circularly-polarized radiation for two HgTe QWs. The advantage of this geometry is the clear separation of the electron ($\mathbf{e}$) and hole ($\mathbf{h}$) resonances, as they are observed for positive and negative external magnetic fields, in agreement with Eq.\,(\ref{eq_drude}). Below $V_g = -6$\,V we observe a hole CR ($\mathbf{h_1}$) in both samples. In the entire region of the applied gate voltage we observe an electronic CR which we denote as $\mathbf{e_1}$. In  the $22$\,nm~sample and at high frequencies ($658$ and $966$~GHz, Fig.\,\ref{fig_spec}(a)), which provide a higher resolution of the cyclotron mass, we observe an additional contribution above $V_g>9 V$, which we identify as carrier type $\mathbf{e_2}$.

Several transmission spectra show contributions in positive fields that are symmetric with respect to electron CR  (see for example Fig.~\ref{fig_spec}(b) at $\sim0.1$~T). We attribute them to mirror peaks of the $\mathbf{e_1}$ CR, a consequence of non-ideally circularly-polarized incident radiation. In some spectra, several anti-symmetrically mirrored peaks are observed that likely correspond to $\mathbf{e_1}$ CR as well  (see Fig.\,\ref{fig_spec}(d) between $\pm0.5$~T). We suggest an admixture of the signal reflected from the backside of the substrate or admixture of the phase signal as possible candidates, the latter effect might be more pronounced for sharper peaks. Full understanding of mechanisms behind these signatures requires further investigations.

The transmission curves can be fitted well using the procedure presented in Sec.\ref{secspec} (solid lines in Fig.\,\ref{fig_spec}). From the analysis of the resonances in the transmission, we obtain the 2D charge density $n_{2D}$, effective cyclotron mass $m_c$, and the scattering time $\tau$ for each carrier type.

Figure~\ref{fig_par} shows the parameters of the CRs. Data obtained at different frequencies overlap well in the case of electrons. On the other hand, holes are characterized by heavier masses, lower carrier density, and mobility. These factors result in much weaker cyclotron signatures in the transmission spectra. The analysis of the hole CRs becomes affected by the noise level, time-related drifts, and other artifacts, resulting in larger fitting errors.  Nevertheless, as seen in Figs.~\ref{fig_par}(d,h), all carrier types demonstrate a linear behavior of the cyclotron frequency ($\omega=2\pi\nu$) with respect to the resonance field ($B_{r}=m_c\omega/e$), satisfying the quasi-classical approach in Eq.~(\ref{eqCR}). Figures \ref{fig_par}(a,e) show a gradual increases of the effective mass of $\mathbf{e_1}$ above $\sim-5$~V reflecting the deviation of the dispersion from parabolic form. At lower voltages, the mass increases with decreasing gate until it stabilizes at around $-8$~V for both samples.

The absolute values of the charge density decrease with the gate voltage for holes and increase for electrons.  Both agree with the sign of the charge carrier types obtained directly from the spectra in Fig.\,\ref{fig_spec}. With the help of the dotted lines in Figs.~\ref{fig_par}(c,g), one can observe a sectional linear relation between the $\mathbf{e_1}$-carrier concentrations and the gate voltage. The slope $\partial n_{2D}/\partial V_g$ of $\mathbf{e_1}$ changes with the emergence of additional carrier type $\mathbf{e_2}$ in the system, since the derivative of total charge density $\partial n_{tot}/\partial V_g$ can be expected to remain constant.  It is evident from Fig.\,\ref{fig_par}(c) that both, the sum of slopes of $\mathbf{e_1}$ and $\mathbf{e_2}$ (above $10$~V) and sum of slopes belonging to $\mathbf{e_1}$ and $\mathbf{h_1}$ (below $-10$~V) match the $\partial n_{2D}/\partial V_g$ of $\mathbf{e_1}$ between $-10$ and $10$~V. Similar holds for the case of the thinner sample (Fig.\,\ref{fig_par}(g)).
The fact that all parameters of $\mathbf{e_1}$ almost stabilize below $-8$~V as in this range the electrons are weakly affected by the gate voltage. We assume this to be a consequence of the Fermi level entering the flat valence band with a high density of hole states.

\subsection{Mass vs. density}
\label{secres_massdens}

\begin{figure}[tbp]
	\includegraphics[width=1\linewidth]{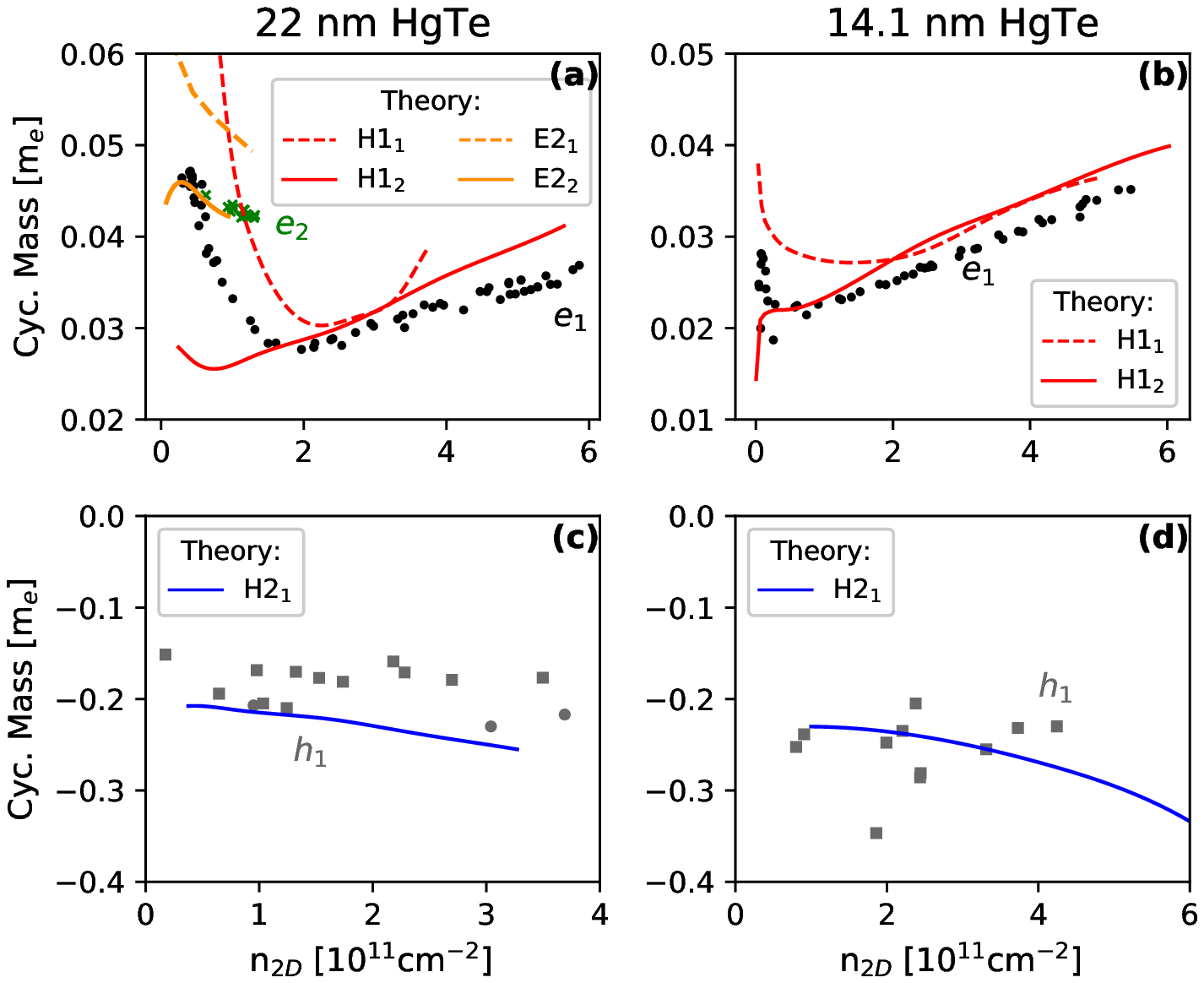}
	\includegraphics[width=1\linewidth]{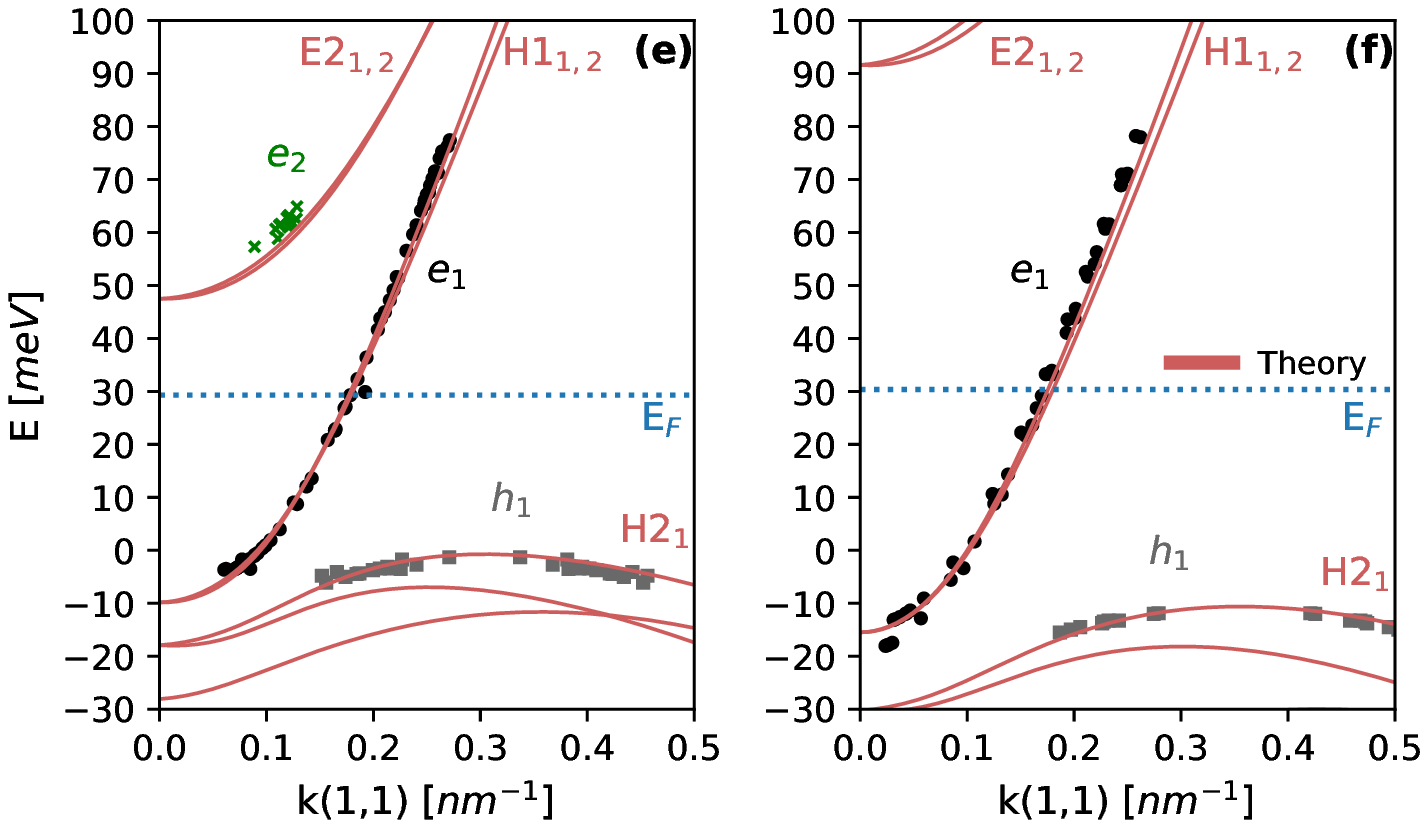}
	\caption{\textit{Comparison of the experimental data with $\mathbf{k \cdot p}$ model calculations.} Contrary to Fig.\,\ref{fig_par},	the cyclotron masses are plotted as a function of density. (a) electron-like and (c) hole-like carrier types corresponding to the $22$~nm sample. (b) electron-like and (d) hole-like carrier types detected in the $14.1$~nm sample.	Full symbols - experimental values, solid lines - theory. (e,f) Band structures of both samples along the (1,1) direction corresponding to $n_{2D}\approx2.5\cdot 10^{11}$~cm$^{-2}$ per each, H1$_1$ and H1$_2$ band. Symbols - experimental data obtained from cyclotron	mass, solid lines are predictions of the $\mathbf{k} \cdot \mathbf{p}$ model.}
	\label{fig_mass}
\end{figure}

Figures~\ref{fig_mass}(a-d) compare the density dependence of the cyclotron mass with predictions of the $\mathbf{k \cdot p}$ model.
While varying the total charge density $n_{tot}$ between $-7\cdot10^{11}$~cm$^{-2}$ and $12\cdot10^{11}$~cm$^{-2}$, we obtained the theoretical values of the electron (hole) charge densities by integrating over the occupied conduction (unoccupied valence) states. Knowing the positions of the Fermi energy, the theoretical values of the cyclotron mass were determined using the quasi-classical Eq.\,(\ref{eqCR}).

The reader should note that data presented in Figures~\ref{fig_mass}(a-d) do not rely on the isotropic approximation argued in Section~\ref{secband}. This representation of data has an additional advantage.  Since the measurements took place in different cooling cycles, we cannot fully rely on the gate voltage representing a good absolute metric of the electronic state of the samples. Moreover, we observe that applying high gate voltages (above $\sim 10$\,V) to the sample lead to the saturation of the total charge density and result in a shift of gate value at which the charge neutrality point is observed, i.e., the previous correspondence between the applied gate voltage and the electronic state of the sample becomes obsolete.

Experimental and theoretical data in Fig.~\ref{fig_mass}(a,b) correspond to electron-like carrier types for both samples. The $\mathbf{k \cdot p}$ theory predicts two spin-split states H1$_1$ and H1$_2$. The mass of H1$_1$ shows signs of divergence at low densities, while the mass corresponding to the H1$_2$ subband stays at lower values and even drastically decreases in the $14.1$\,nm sample. This separation occurs due to an inversion asymmetry which results in energetic minima at very small finite $k$-values of H1-bands. Similar behavior is also observed for the spin-split states E2$_1$ and E2$_2$. Although the theory predicts all electronic states to be single-degenerate with $D=1$, it appears that the resolution of the CR experiments did not allow to observe the spin-splitting of electronic states leading to degeneracy $D=2$. In order to be able to compare theory and experiment on the same graph, we  divide  the experimental densities of carriers types $\mathbf{e_1}$ and $\mathbf{e_2}$ by $D=2$ in Fig.~\ref{fig_mass}(a,b).

The comparison between the experimental points with theory allows us to recognize the carrier type $\mathbf{e_1}$ as the fingerprint of the first conduction H1 band since the corresponding data seem to match with the mean value of H1$_1$ and H1$_2$. We note that the splitting of these two spin-polarized states was observed at higher densities with the SdH analysis of the $14.1$\,nm sample, which is given in the Supplementary in Ref.~\cite{supp}.

In the case of $22$\,nm sample, the data suggest that the charge carrier type $\mathbf{e_2}$ is linked with the second conduction band E2. Carrier type $\mathbf{e_2}$ cannot be linked to an individual spin-polarized subband due to the measured double degeneracy of the mode.

As shown in Figs.\,\ref{fig_mass}(c,d), the carrier type $\mathbf{h_1}$ corresponds to the first valence bands (H2$_1$) in the systems, i.e., the spin-polarized hole band with the two symmetrical islands in the $(k_x,k_y)$-plane of the band structure. Similar to previous results~\cite{Minkov2017}, the theoretical hole mass slightly decreases with the carrier concentration. This trend is supported by the scattering of the experimental points for the $22$\,nm sample, however, data of the thinner sample do not provide the accuracy for such a conclusion.

\subsection{Band structure of HgTe 2D semimetal}
\label{secres_bs}

According to the identification of the carrier types in the previous section and in the SdH data, we assume the spin degeneracy ($D=2$) for $\mathbf{e_1}$ in both samples and $D=2$ for $\mathbf{h_1}$. The position of the experimentally reconstructed valence band maximum was shifted to the value provided by the theoretical model. As pointed out in Section~\ref{secband}, for hole-like carrier types we calculate the $k$-vector along the (1,1) direction as $k= \sqrt{2}k_0 \pm k_{eff}$ with $k_{eff}=\sqrt{2 \pi n}$.
In  case of the $22$\,nm sample, the carrier type $\mathbf{e_2}$ is assumed to be spin-degenerate with $D=2$.

The experimental band dispersions, calculated as described in Sec.~\ref{secband}, are shown in Fig.\,\ref{fig_mass}(e,f) as solid symbols. Direct integration lacks in providing the absolute energy position of the bands. Since we assume that the gate voltage defines a constant Fermi level in the film, the bands are vertically aligned to each other by referring to the gate voltage at which they were mutually detected.

As discussed in Sec.~\ref{secmod}, varying the total charge density $n_{tot}$ by gate voltage in an asymmetric system results in a variation of the corresponding band structure. While these effects play a major role in thicker samples~\cite{Gospodaric2020}, in our case, the differences between band structures calculated at boundary values of $n_{tot}$ are not significant with the accessible energy resolution. Therefore the experimental results are compared to a single theoretical band structure, that was obtained at a total density $n_{tot}=5\cdot10^{11}$~cm$^{-2}$, i.e. $n_{2D}\approx2.5\cdot 10^{11}$~cm$^{-2}$ per each band, H1$_1$ and H1$_2$, with the Fermi levels as indicated in Fig.\,\ref{fig_mass}(e,f). According to the theory estimates for the (0,1) direction, the first valence subband lies $\sim 5$\,meV lower than along the (1,1). Although seeing these maxima in the cyclotron analysis would be very interesting, we cannot reach them in the present experiment. Therefore, we show the (1,1) cut only, where the valence band is reached.

As clearly seen, the first conduction band H1 is formed by the carrier type $\mathbf{e_1}$ in both samples. The experimental resolution of the CR analysis did not allow to observe the spin-splitting of the H1 band.
We note that the steeper slope of the experimental points $\mathbf{e_1}$ at higher energies is probably a result of stronger structure-inversion asymmetry at high gate voltages~\cite{Gospodaric2020}. At low energies, the carrier type $\mathbf{h_1}$ nicely overlaps with the valence band H2$_1$, which are spin-polarized and double-valley degenerate. 
The $\mathbf{e_2}$, detected in the $22$\,nm sample, correspond to the E2 band and, similar to $\mathbf{e_1}$, the experiment showed the spin-degenerate character of this band.

\section{Conclusions}

We described the results of the cyclotron resonance in two semimetallic HgTe samples with thicknesses of $14.1$ and $22$~nm in the sub-THz frequency range. The quasi-classical approach is utilized to reconstruct the parameters of the charge carriers, which is approved by the linearity of the cyclotron frequency in external magnetic fields. With the help of $\mathbf{k\cdot p}$ models of the electronic configurations of both samples, several CR modes are recognized as fingerprints of the first valence band, first and second conduction bands. The results show a good overlap with the predictions of the $\mathbf{k \cdot p}$ theory. The outcome of the CR analysis was also supported by the Shubnikov-de Haas analysis of the field-dependent capacitance measurements. Experimentally obtained band structure showed an overlap of $\sim 5-10$\,meV between the first conduction and first valence band, in turn confirming the existence of the semimetallic state in HgTe QWs in this range of thicknesses. Furthermore, the results confirm the anomalous two-fold valley degeneracy of the first hole-like valence band.

\section*{Acknowledgments}

We acknowledge valuable discussion with G. M. Minkov. This work was
supported
by Austrian Science Funds (Grants No. W-1243,  I3456-N27, TAI 334-N ),
by Russian Foundation for Basic Research (Grant No. 17-52-14007),
and by German Research Foundation Grants No. AS 327/5-1 and SFB 1143
(project-id 247310070).

\section{APPENDIX A: MATERIALS AND METHODS}
\label{secexp}

\subsection{Samples}

Two CdHgTe/HgTe/CdHgTe quantum wells with $14.1$ and $22$\,nm thick HgTe layers were grown by molecular beam epitaxy on {GaAs} substrates~\cite{mikhailov_jnanotech_2006, kvon_ltp_2009}  with (013) surface orientation. A $6$~$\mu$m-thick CdTe buffer layer between the layered structure and the substrate ensured that the lattice variation was not abruptly changing, which resulted in high electron mobilities $\mu \sim 2 \cdot 10^5$\,cm$^2/$Vs. For the optical experiments, the $5\times5$~mm plates were cut from the wafer.  On top, the structure was covered by a multilayered insulator SiO$_{2}/$Si$_3$N$_4$ and a metallic Ti-Au layer, which acted as a semitransparent gate electrode. The gate was grown in a cross-like shape in order to cover the center of the sample for THz transmission measurements and to allow four electrical contacts on the corners of the sample for simultaneous transport measurements.


\subsection{Technique}

The CR response in the investigated systems was studied using magneto-optical transmission technique with controlled polarization of light~\cite{shuvaev_sst_2012, dziom_2d_2017}. Backward-wave oscillators were employed to produce continuous monochromatic light in the frequency range $100$\,GHz\,--\,$1000$\,GHz. The polarization of the incident radiation was set to be circular, which in turn allows observing either electron or hole CRs, depending on the sign of the applied magnetic field. An external magnetic field was provided by a split-coil superconducting magnet and was applied parallel to the $\mathbf{k}$-vector of the terahertz radiation, i.e., in the Faraday geometry.
The experiments were conducted with a sweeping magnetic field at various fixed frequencies of incident radiation. In addition, frequency-dependent spectra in zero magnetic field were measured. All experiments were done at the lowest temperature of our spectrometer $T=1.8$\,K.

\subsection{Cyclotron resonance}
\label{secspec}

For circularly-polarized radiation the Drude model for dynamical conductivity in the quasi-classical approximation~\cite{palik_rpp_1970, Gospodaric2020} can be written as:
\begin{equation}
\sigma_{+} =\sum_{j} \frac{\sigma_{0,j}}{1- i\tau_j (\omega + \Omega_{c,j})} \ ,
\label{eq_drude}
\end{equation}
where for each active carrier type $\sigma_{0,j}=n_j e^2\tau_j/m_{c,j}$ is the two-dimensional DC-conductivity, $\Omega_{c,j}=eB/m_{c,j}$ is the resonance frequency, $n_j$ is the two-dimensional density, $\tau_j$ is the
intrinsic scattering time, $m_{c,j}$ is the effective cyclotron mass, and $e$ is the electron charge. The transmission of circularly-polarized radiation through a metallic film with a single charge carrier and a thickness of $d\ll\lambda$ can be calculated as:
\begin{equation}
t_+ =1- \frac{i}{\tau_{\mathrm{SR}}}\frac{1}{(\omega+i \Gamma)-\Omega_c} \ .
\label{eqtr}
\end{equation}
Here, $\Gamma = {1}/{\tau}+ {1}/{\tau_{\mathrm{SR}}}$ corresponds to the "total" scattering rate,
$1/\tau$ is the transport scattering rate, $1/\tau_{\mathrm{SR}} = ne^2 Z_0/2m_c$ is the superradiant damping~\cite{gospodaric_prb_2019}, and $Z_0$ is the impedance of the free
space. Eq.~(\ref{eqtr}) serves as a convenient demonstration that a resonant response of the system with a single charge carrier is expected in transmission spectra. However, for a proper analysis of the experimental data, we  employ a model with multiple carrier types in the system, which also takes into account the reflections inside the substrate, The procedure utilizes similar algebra as described previously~\cite{shuvaev_prl_2011, dziom_ncomm_2017, Gospodaric2020}.

\subsection{Reconstruction of the band structure}
\label{secband}

In the quasi-classical approximation transitions between several Landau levels overlap and the cyclotron frequency $\Omega_c$ can be written in terms of the cyclotron mass as~\cite{ashcroft_book}:
\begin{equation}
m_c \equiv \frac{eB_r}{\Omega_c} =
\frac{\hbar^2}{2 \pi }\frac{\partial A}{\partial E} \Big|_{E=E_F} \ .
\label{eqCR}
\end{equation}
Here $B_r$ is the resonance magnetic field, $A$ is the area in the reciprocal space enclosed by the contour of the constant energy $E$, and $E_F$ is the Fermi energy. If the Fermi area is magnetic field-independent at least at low fields, Eq.\,(\ref{eqCR}) leads to a linear relation between $\Omega_c$ and $B_r$ independently on the form of the dispersion relations.
To compare theory and experiment without relying on the isotropic approximation, the cyclotron mass can be plotted directly as a function of the 2D density of each individual carrier type~\cite{Gospodaric2020}.

In further approximation for 2D materials, the local isotropy of the bands can be assumed. This leads to a simple relation between the Fermi-vector $k_F$ and the Fermi-area: $A=\pi k_F^2$. Then Eq.\,(\ref{eqCR}) can be rewritten as:
\begin{equation}
\frac{\partial E}{\partial k} \Big|_{E=E_F} = \frac{\hbar^2 k_F}{m_c}\ ,
\label{eqCR2}
\end{equation}
and thus can be directly integrated to obtain $E(k)$.

We note however that especially for the hole islands the isotropic approximation does not hold.
Nevertheless, at lower hole concentrations the islands can be approximated as
circles (see Fig.~\ref{fig_th}(b)) with an effective radius $k_{eff}$ related to the
Fermi-surface area as $A=\pi k_{eff}^2$.
Of course, the exact relation between $A$ and $k_{eff}$ can be
calculated from the theory. We believe, however, that a reasonable picture
of the band structure can be obtained within this approximation as well.
A direct comparison between theory and experiment can be done using
an approximation-independent plot of cyclotron masses vs. density. This presentation is not sensitive to
approximations done in Eq.\,(\ref{eqCR2}).

To calculate the experimental band structure via Eq.\,\ref{eqCR2}, the charge density of electrons is transferred to the electron momentum using the relation $k=\sqrt{4\pi n/D}$ with $D$ being the degeneracy of the band. In case the Fermi area is shifted to a point $\mathbf{k_0}$ in the Brillouin zone, the wavevector in Eq.\,(\ref{eqCR2}) is calculated as $\mathbf{k_{eff}}=\mathbf{k}-\mathbf{k_0}$. Here $\mathbf{k_0}$ has to be taken from, e.g., model calculations. In a further improvement, deviations of the Fermi surface from the circle can be taken into account. However, these details are beyond the accuracy of the present experiment.

\subsection{Theoretical band structure}
\label{secmod}

The band structure of the strained HgTe QWs has been calculated using the eight-band $\mathbf{k \cdot p}$ model in an envelope function approach which includes the coupling between the lowest conduction band $\Gamma_6^c$ and the topmost valence bands $\Gamma_8^v$ and $\Gamma_7^v$. Further details about the model can be found in Ref~\cite{novik_prb_2005} where specific $\mathbf{k \cdot p}$ parameters are given.
Assuming that HgTe QWs are grown on a CdTe substrate, strain effects due to the lattice mismatch between HgTe and CdTe were taken into account applying the Bir-Pikus formalism~\cite{bir_pikus_1974}. A generalization of the $\mathbf{k \cdot p}$ model for structures grown on high-index-planes \cite{los1996generalization} has been used to include additional terms into the Hamiltonian that are responsible for coupling of states for the (013) growth direction.

The calculations have been done taking into account structure and bulk inversion asymmetry~\cite{winkler2003spin}. Whereas in the experiment the carrier density in the QW is tuned by the gate voltage, in the model the variation of the doping in the top barrier is assumed, while the doping in the barrier on the substrate side is taken to be constant. Asymmetric barrier doping results in the asymmetric distribution of the Hartree potential which has been determined by solving self-consistently the eigenvalue problem and the Poisson equation for the two-dimensional charge carriers in the QW \cite{novik_prb_2005}. Bulk inversion asymmetry of the zinc-blende crystal structure gives rise to the Dresselhaus spin-orbit interaction. Here, the bulk-inversion asymmetry terms were linear in momentum for the valence bands $\Gamma_8^v$ and $\Gamma_7^v$, while the contribution from the coupling between the conduction and valence bands was quadratic in momentum  \cite{winkler2003spin}. The following bulk inversion asymmetry parameters taken from Refs.~\cite{winkler2003spin,cardona1986terms,winkler2012robust} were used: $C_k(HgTe)=-7.46$~meV nm, $B^+_{8v}(HgTe)=-200$~meV nm$^2$, $B^-_{8v}(HgTe)=10$~meV nm$^2$, $B_{7v}(HgTe)=-200$~meV nm$^2$, $C_{k}(CdTe)=-2.34$~meV nm, $B^+_{8v}(CdTe)=-224.1$~meV nm$^2$, $B^-_{8v}(CdTe)=-6.347$~meV nm$^2$, $B_{7v}(CdTe)=-204.7$~meV nm$^2$.

%

\end{document}